\documentstyle[psfig]{l-aa}     

\let\ni=\noindent

\begin{document}

   \thesaurus{(12.03.3;  
               12.04.2;  
               13.25.2;  
               13.25.3)  
             }

\title {The large scale structure of the soft X-ray background.
II: Galaxies}
\author{Andrzej M. So\l tan\inst{1,2}, G\"unther Hasinger\inst{3},
Roland Egger\inst{2}, Steve Snowden\inst{2,4,5}  and Joachim Tr\"umper\inst{2}}

\offprints{A.\,M.\,So\l tan, Copernicus Center address}

\institute{$^1$~Copernicus Astronomical Center, Bartycka 18, 00-716 Warsaw,
                Poland \\
           $^2$~Max Planck Institut f\"ur extraterrestrische Physik,
                D-85748 Garching bei M\"unchen, Germany \\
           $^3$~Astrophysikalisches Institut, D-14482 Potsdam, Germany \\
           $^4$~NASA/GSFC, LHEA Code 666, Greenbelt, MD\ \ 20771, USA \\
           $^5$ Universities Space Research Association}

\date{Received May 1996; accepted }

   \maketitle
   \markboth{A.\,M.\,So\l tan et al.: The large scale structure of the
soft X-ray background.\,II}{A.\,M.\,So\l tan et al.: The large scale structure
of the soft X-ray background.\,II}

   \begin{abstract}
The intensity of the soft X-ray background is correlated 
with the distribution of galaxies. To demonstrate this, magnitude limited 
galaxy samples extracted from bright galaxy catalogues and the Lick counts are
utilized. Significant correlations are detected for all magnitude
ranges, i.e. $10 < m_{\rm B} \la 18.6$. The local X-ray volume emissivity
between 0.5 and 2.0 keV correlated spatially with the galaxy population
falls in the range
($8\times 10^{38}$ -- $1\times 10^{39}$)\,erg\,s$^{-1}$\,Mpc$^{-3}$
($H_{\circ}=100$\,km\,s$^{-1}$\,Mpc$^{-1}$). Without evolution, this
could account for 30\,\% -- 40\,\% of the total X-ray background, an
amount consistent with analogous estimates based on X-ray experiments
at higher energies. The comparison of correlation amplitudes of the
X-ray background with galaxies at different apparent magnitudes
indicates that roughly half of the emission correlated with galaxies
can be produced in extended regions substantially larger than optical
extent of a typical galaxy. A crude estimate for the average size of
these sources is $\sim 1$\,Mpc.  This extended signal in various
correlation functions is possibly produced by X-ray emission of hot gas
in clusters or groups of galaxies, although a contribution from
individual galaxies cannot be ruled out.
   \keywords{Cosmology:observations, diffuse radiation, X-rays:galaxies,
             X-rays:general}
   \end{abstract}
%

\section{Introduction}
\ni At least 60\,\% of the soft extragalactic X-ray background (XRB) has
been resolved into discrete sources (Hasinger et al. 1993). Thus, the
very nature of the XRB is now essentially understood. Apart from optical
identificatons and the determination of lumnosity functions
of the contributors to the XRB, many current investigations now
concentrate on more subtle properties, such as surface brightness
fluctuations at various angular scales.
Below $\sim 1^{\prime}$ the inhomogeneity of the XRB is directly
related to individual sources. An investigation of the statistical character
of these fluctuations shows that at least another 15\,\% of the total
XRB is produced by discrete sources (e.g. Hasinger et al. 1993). At
intermediate scales, $1^{\prime}$ to $\sim 30^{\prime}$, deviations from
a smooth distribution are generated primarily by rich clusters of galaxies
(e.g. Briel \& Henry 
1993, and references therein). Fluctuations at these angular scales could
potentially also result from the nonuniform distribution of sources
contributing to the XRB. Upper limits on the amplitude of the XRB
variations which impose significant
constraints on the clustering properties of sources
have been obtained by several authors (e.g. So\l tan \&
Hasinger 1994, and references therein).

At still larger scales, $\theta \ga 0.\!\!^{\circ}5$, low amplitude
inhomogeneities could originate from extended nearby sources and/or
from sources substantially larger than typical rich clusters.
The structure of the XRB at scales above $0.\!\!^{\circ}5$ has been
investigated in the first paper of this series (So\l tan et al. 1996,
hereafter Paper I). In that work we have measured total fluctuations
of the XRB using
the auto-correlation function (ACF) of the {\em ROSAT} All-Sky Survey (RASS)
maps (see below). A positive ACF amplitude has been detected for 
separations up to $\theta \la 6^{\circ}$. The measurement of the
cross-correlation function between the XRB maps and the distribution of
Abell clusters revealed that about $1/3$ of the ACF amplitude of the XRB
at separations $\la 4^{\circ}$ can be accounted for by a new class
of X-ray emitters 
associated with rich clusters of galaxies. The origin of the remaining XRB
fluctuations is at present unknown. It has been suggested that nearby
poor groups of galaxies could potentially contribute to the XRB
fluctuations (Hasinger 1992).

This is a second paper of the series which presents further results of
our study of the large scale fluctuations of the soft XRB. The aim of
this investigation is to determine the average X-ray emissivity of normal
galaxies using the extensive RASS data. The X-ray properties of a large number
of nearby individual galaxies have been examined on many occasions
(e.g. Fabbiano 1989, Eskride et al. 1995, and references therein), but this
type of observation cannot be effectively used to determine the total
local volume emissivity associated with the overall galaxy population.
The main thrust of the present investigation is to utilize correlations
between distributions of galaxies and the RASS maps.  This was recognized 
as a potential method for studying the XRB several years ago by Turner \& 
Geller (1980). They obtained an upper limit of
$\sim 50$\,\% on the fraction of the XRB in the {\em Uhuru} energy band
($2 - 6$\,keV) produced by any class of sources ``represented among
bright (m$_{\rm pg} \leq 15.5$) galaxies''. Jahoda et al. (1991, 1992)
calculated the cross-correlation function (CCF) between the {\em HEAO-1} A-2
all-sky survey ($2 - 10$\,keV) and the galaxy surface density based on
samples from the Uppsala General Catalogue (Nilson 1973, hereafter UGC)
and the European Southern Observatory catalog (Lauberts 1982).
The detection of a positive correlation signal led Jahoda et al. to the 
conclusion
that non-evolving sources can produce as much as 50\,\% $\pm$ 30\,\% and
70\,\% $\pm$ 40\,\% of the XRB respectively, using the two galaxy catalogues.
Their figures were revised downward to ($30 \pm
15$)\,\% by Lahav et al. (1993) who analysed the cross-correlation of
the {\em Ginga} data with the UGC and IRAS galaxy samples. The reason
for this reduction resides in different assumptions about the spatial
distribution of
galaxies contributing to the XRB. Jahoda et al. did not take into
account the clustering of galaxies and thus overestimated the average 
ratio of X-ray
emissivity to the surface density of galaxies. Lahav et al. have shown
that clustering effects are essential in this case.

A further refinement of this method was achieved by Miyaji et al.
(1994), who investigated the cross-correlation between {\em HEAO 1} A-2
XRB maps and samples of galaxies selected from the {\em IRAS} point
source catalogue. To model the CCF of the XRB -- galaxy distribution,
they derive formulae describing clustering effects and correlations of
X-ray and IR luminosities of the sources under examination. Miyaji et al.
point out that their estimates of the total X-ray volume emissivity
above 2\,keV are model-dependent. Analogous calculations have been
performed by Carrera et al. (1995) using {\em Ginga} scans in the $4 -
12$\,keV band. They found that $\la 10 - 30$\,\% of the XRB could be
produced by a non-evolving population of galaxies.

The angular resolution of both the {\em HEAO 1} A-2 and {\em Ginga} LAC
experiments
is defined by the collimator sizes, $1.\!\!^{\circ}5 \times 3^{\circ}$
and $1^{\circ} \times 2^{\circ}$, respectively. All the cosmic signal of
correlations on scales $\la 1^{\circ}$ is hidden behind the strong
correlation amplitude generated by the poor instrumental angular resolution.
Because typical correlations between the XRB and galaxy distributions at
scales $\ga 1^{\circ}$ are very weak (see below), investigations based on
non-imaging optics are virtually restricted to zero-lag cross-correlation
measurements. For instance, this is illustrated by Fig. 2 of Carrera et al.
(1995). The correlations are determined over a wide range of separations,
but above $\sim 1^{\circ}$ any CCF signal is lost in the noise.

Three deep {\em ROSAT} pointings have been used by Roche et al. (1995) to
determine the CCF of the unresolved XRB and faint (B $< 23$) galaxies. The 
angular resolution of the X-ray telescope allowed for measurements of the
CCF amplitude at several separations. Although the scatter between fields is
substantial, the mean CCF clearly exhibits an extension exceeding the width of
the Point Spread Function. The X-ray emissivity estimated by Roche et al.
converted to $H_{\circ} = 100$\,km\,s$^{-1}$\,Mpc$^{-1}$
amounts to $(10.6 \pm 1.4)\times 10^{38}$\,erg\,s$^{-1}$\,Mpc$^{-3}$ in the
0.5 -- 2\,keV band. Treyer \& Lahav (1995) applied a correlation formalism to 
investigate the relationship between the population of faint
blue galaxies (B = 18 -- 23) and the soft XRB. They carefully reevaluated
the Roche et al. results and estimate the comoving volume emissivity at
$(6 - 9) \times 10^{38}$\,erg\,s$^{-1}$Mpc$^{-3}$ in the $0.5 - 2.0$\,keV
band.

Total X-ray emissivity correlated with the galaxy distribution is used
to estimate contribution of ``normal'' objects to the total XRB.
Term ``normal'' is commonly used to the overall galaxy population as distinct
from variety of objects showing any form of activity. 
In the present paper, normal objects are those which in the local Universe
follow the distribution of sample galaxies.
Our analysis cannot be used to distinguish between various
mechanisms of galaxy emission. In particular, ``classical'' normal galaxy
emission due to X-ray binaries and SNRs would produce a similar correlation
signal to that generated by nuclear galaxy activity distributed among the
population of otherwise ``normal'' galaxies (e.g. Elvis et al. 1984). One
could separate stellar signal from scaled down AGN behaviour if the spatial
distributions of both types of galaxies are different. Several
investigations (Iovino and Shaver 1988, Kruszewski 1988, Andreani and
Cristiani 1992, Mo and Fang 1993) show that a class of powerful AGNs, viz.
quasars exhibit in fact stronger clustering than general population of
galaxies. However, clustering properties of AGNs at the faint end of
the nuclear luminosity function are not known and it is likely that
the spatial distribution of these galactic nuclei is identic to the general
galaxy distribution.

Identification of X-ray sources directly show what kinds of objects
contribute to the XRB at various flux levels. Here we summarize basic
data on this subject for comparison with the present results.
The largest contribution to the XRB above {\em ROSAT} sensitivity
threshold comes from AGNs (e.g. Boyle et al. 1993 and references therein).
The deepest {\em ROSAT} exposure in the {\em Lockman Hole} resolves
about 60\,\% of the background with AGNs clearly seen in high proportion
(Hasinger et al. 1993, Bower et al. 1996). Systematic identifications
of X-ray sources in a large number of {\em EINSTEIN} and {\em ROSAT} fields
show that QSOs with luminosities in the range $10^{42}$ -- $10^{46}$
erg\,s$^{-1}$ in the soft X-rays produce 30\,\% to 90\,\% of the XRB at
2\,keV. Although, due to poor estimates in the faint end
and high redshift evolution of the X-ray luminosity function, above limits
are still wide, it is well established that the AGN contribution to the XRB
is substantially larger than the contribution of ordinary galaxies, i.e.
galaxies which produce X-rays through standard thermal processes.

Detailed studies of nearby galaxies and the Milky Way reveal
a variety of X-ray sources associated with several galaxy components.
The total X-ray emission of an individual galaxy is a mixture of several
constituents. The integrated flux produced by single and binary stars,
supernova remnants, thermal emission by hot gas and -- in some objects
-- non-thermal radiation produced in an active nucleus create a
complex X-ray map for each galaxy (e.g. Fabbiano 1989 and references
therein). Observations using the {\em EINSTEIN} and {\em ROSAT} satellites
show that normal galaxies of all morphological types are spatially
extended X-ray sources with luminosities in the range of $\sim 10^{38}$
to $\sim 10^{42}$\,erg\,s$^{-1}$. Separate from this ``normal''
X-ray emission, some fraction of galaxies exhibit non-thermal nuclear activity.
It is likely that luminosities of active galactic nuclei extend to
arbitrary low levels and the distinction between normal and active galaxy
could in some cases be a matter of convention (e.g. Elvis et al. 1984).
Using ratio of X-ray to optical luminosities (excluding non-thermal
nuclear component) integrated emission of normal galaxies
contributes $\sim 13$\,\% to the XRB at 2\,keV (Fabiano 1989).

Clusters of galaxies constitute a separate class of X-ray sources.
Observations of distant clusters (e.g. Edge et al. 1990, Henry et al. 1992,
Ebeling 1993, but see also Ebeling et al. 1996) indicate -- although
this question still could be debated --
that their X-ray luminosity function undergoes
strong evolution in the sense that local volume emissivity is greater than that
at high redshifts. According to various estimates, contribution of rich,
Abell-type, clusters to the XRB falls between 5\,\% and 10\,\%.

Correlations between the XRB and selected galaxies provide information
on all kinds of X-ray sources which are spatailly correlated with
those galaxies. The CCF signal represents integrated emission of
sources occupying specific volume. Luminosity density calculated
this way comes both from normal galaxies and AGNs as well as from
clusters of galaxies (see below). The present calculations give
the X-ray volume emissivity but do not allow to isolate individual objects.
On the other hand, identifications of sources in flux limited
samples provide information on discrete sources, but are ineffective
method to calculate total volume emissivity.

In this context it is important to note that our samples were constructed
using apparent magnitudes. Any X-ray emission which is not correlated with
galaxies (e.g. hypothetical X-ray sources associated with dwarf galaxies
in voids) is not taken into account. Thus our measurement is actually
a lower limit for the total X-ray emissivity if other kinds of X-ray
sources are common in the local Universe. 

The basic framework of the present investigation is analogous to the
work by Roche et al. (1995), although we are using totally different
observational material both in the X-ray and optical domains.  The main
advantage of the present analysis resides in the high quality of the
X-ray data accumulated in the RASS. The massive amount of the RASS data
warrants not just a quantitative improvement of the measurement
accuracy, but also a substantial extension of the scope of analysis as compared
to the previous investigations. Our objective is to determine the CCFs
between the XRB and several magnitude limited galaxy samples. These
observed CCFs are compared with predictions based on the correlations of
galaxies measured in each sample separately and between the samples. Then,
we estimate the ratio of the X-ray--to--optical volume emissivities associated
with galaxies and -- using optical luminosity density as normalization
-- we calculate the X-ray volume emissivity. The organization of the paper
is as follows: the X-ray and optical material used in the investigations is
described in Sections 2 and 3, respectively. Procedures used to
determine the CCFs and their uncertainties are presented in Section
4. An analysis of the observed CCFs and the construction of models which
properly reproduce the observations is given in the Section 5. We examine
some properties of the X-ray emission in Section 6 and conclude
our investigation with a short discussion in Section 7.

\section{Selection of the X-ray data}

\ni The {\em ROSAT} (Tr\"umper 1983) All-Sky Survey with the PSPC
(Pfeffermann et al. 1987) is used in the analysis.
For a comprehensive description of the RASS see Snowden \&
Schmitt (1990) and Voges (1992). Basic references and relevant
characteristics of the RASS are given in Paper I. Here we summarize
only the essential information. Various effects and constituents
contaminating the cosmic signal (particle background, solar scattered
X-rays and ``short- and long-term enhancements'') have been extracted
in a complex and laborious procedure as described by Snowden et al.
(1995). Within the {\em ROSAT} energy band
($0.1 - 2.4$\, keV), the amplitude of the galactic component relative
to the extragalactic signal increases drastically towards soft energies.  
In Paper I we discuss this question in some detail and find that useful
information on the fluctuations of the extragalactic XRB component are
concentrated mainly in two energy bands in the hard portion of the RASS
labelled R5 and R6. The band R5 is centered at $\sim 0.8$\,keV and in terms
of puls-height invariant (PI) channels includes channels $70-90$,
while the band R6 is defined by channels $91-131$ and is centered at $\sim
1.1$\,keV (Snowden et al. 1994b). Although the bands R5 and R6 cover
overlapping energy ranges, they differ strongly in the level of the
galactic contribution. It is shown in Paper I that a specific linear
combination of the count rates in bands R6 and R5 is significantly less
contaminated by soft emission from hot plasma in the Galaxy than each
of those bands separately. Because in the present investigation we
are interested in the extragalactic component of the XRB, we use exactly
the same procedure as in Paper I to obtain data free from the local
contamination, viz. we utilize a region in the North Galactic Hemisphere:
\begin{equation}
70^{\circ} < l < 250^{\circ}, \;\; b > 40^{\circ},
\end{equation}
in which the count rate of the extragalactic component in the band R6,
$C\!R_{R6}^{ext}$, is defined as
\begin{equation}
C\!R_{R6}^{ext} = 1.15\,C\!R_{R6} - 0.23\,C\!R_{R5},
\end{equation}
where $C\!R_{R5}$ and $C\!R_{R6}$ denote the count rates in the respective
bands (see Paper I for details).

For the purpose of the present paper, the RASS maps are represented by
an array of count rates in pixels. Pixels of
$12^{\prime} \times 12^{\prime}$ are used in our calculation. This size
coincides roughly with the area containing 90\,\% of the counts
produced by a point source.

\section{Optical data -- galaxy distribution}

\ni Galaxy data were divided into several magnitude limited samples.
To construct maps of the galaxy distribution suitable for further
analysis, we have used the {\em Catalogue of Principal Galaxies},
hereinafter, PGC described by Paturel et al. (1989) and Shane and
Wirtanen (1967, hereafter SW) counts in $10^{\prime}$ pixels kindly
provided to us in electronic form by Dr. M. Kurtz. Galaxies
selected from the PGC are divided into four samples according to
apparent magnitude limits: sample 1 contains galaxies with $10 < m <
12$, sample 2 with $12 < m < 14$, sample 3 with $14 < m < 15$ and sample 4 
with $15 < m < 16$. The first three samples are statistically complete, while in
sample 4 some galaxies at the faint magnitude limit are missing due to
the incompleteness of the PGC (see below). The fifth sample contains
galaxies from the SW counts excluding those present in the first four
samples. Thus, it comprises the weakest and -- on the average -- the
most distant galaxies.

To assess completeness of sample 4, we have used the slope of the
number--magnitude relation. In the relevant magnitude
range it is approximately equal to 0.55 (Driver et al. 1994), which is
slightly below the slope expected in the Euclidean non-expanding model
of 0.6. The number of galaxies expected in sample 4, based on the
extrapolation from samples 2 and 3 and a slope of 0.55, is 6\,\% larger
than that actually listed. Using the Euclidean slope of 0.6, this difference 
increases to 23\,\%.

The average number of galaxies in $12^{\prime}\times 12^{\prime}$ pixel in the
sample 5 is 1.730. We derive the approximate faint magnitude limit in this
sample of 18.6\,mag, using the bright end limit of 16\,mag and assuming the
slope of the relation between number counts and apparent magnitude $\approx
0.5$ (Driver et al. 1994). A summary of the relevant data on the galaxy samples
is given in Table 1.

\begin{table*}
\caption[ ]{Galaxy sample parameters}
\vspace*{0.10cm}
\begin{tabular}{ccclcl}
\hline
Sample&\multicolumn{3}{c}{Mag range~~}&$\rho \;\;[\mbox{pxl}^{-1}]$&
                                     $z_{\mbox{mean}}$\\
  1   & 10 & -- & 12   & $1.26\times 10^{-3}$ & 0.004\\
  2   & 12 & -- & 14   & $6.60\times 10^{-3}$ & 0.010\\
  3   & 14 & -- & 15   & $1.80\times 10^{-2}$ & 0.018\\
  4   & 15 & -- & 16   & $6.05\times 10^{-2}$ & 0.024\\
  5   & 16 & -- & 18.6 & $1.730$              & 0.062\\
\hline
\end{tabular}
\end{table*}

\ni In the last column of Table 1 we list the mean galaxy
redshift in each sample. For samples 1 through 4 $z_{\mbox{mean}}$ denotes
the average value calculated for galaxies with known redshifts. In sample 1
all galaxies but one have measured redshift. The fraction of such galaxies
in samples 2, 3 and 4 drops to 0.91, 0.55 and 0.14, respectively. The
$z_{\mbox{mean}}$ in sample 5 is obtained from the model calculations
in Section 5.2.

\section{Cross-correlation of X-ray maps with galaxy distributions}

\ni The wide range of X-ray luminosities of galaxies and the heterogeneity of 
spatial structures and emission mechanisms complicate estimates of average
sample properties. The RASS offers a possibility to measure the overall galaxy
emission using a large sample of galaxies. Obviously, the angular resolution
and sensitivity of the RASS is not adequate to study weak objects individually.
The typical signal-to-noise ratio for the X-ray detection of
galaxies is usually less than one and most galaxies cannot be
recognized as distinct sources. Nevertheless, X-ray fluctuations
produced even by the most distant galaxies in our samples (see below)
are easily measured using the correlation technique. This allows us to measure 
the average X-ray emission associated with the overall galaxy population.

Rich galaxy clusters constitute a well established and recognized
class of sources in which the X-ray emission originates in the hot intracluster
gas and is not linked with individual galaxies. With the exception of the few
brightest clusters, the flux produced in individual galaxies cannot be
separated from the cluster emission in the RASS maps. Thus, the 
question of the cluster contribution to the local luminosity density
associated with galaxies could not be addressed in the present
investigation. Furthermore, the distribution of galaxies is correlated
with the distribution of rich clusters of galaxies over a wide range in
separation (Seldner \& Peebles 1977). This correlation also affects our
present analysis. The effects produced by Abell clusters are measured
directly in Section 4.2 where we obtain the CCFs of the RASS maps and
galaxy samples using two sets of data. In the first case, the full
observational material including Abell clusters is utilized. Then we remove
from the data areas containing Abell clusters and repeat the CCF calculations.

The contribution of clusters and groups not included in the Abell catalogue is
discussed in the Section 6. 

\subsection{Definitions}
\ni The CCF is defined in a standard way:
\begin{equation}
w_{Xg}(\theta) = {\langle \rho_X(\mbox {\bf n}) \rho_g
(\mbox {\bf n}^{\prime})
\rangle_{\theta} \over \langle \rho_X \rangle \langle \rho_g\rangle} - 1,
\end{equation}
where $\rho_X(\mbox{\bf n})$ is the intensity of the X-ray background in
the direction {\bf n} and
$\rho_{g}(\mbox{\bf n}^{\prime})$ is the surface density of galaxies in
the direction {\bf n}$^{\prime}$, $\langle ... \rangle$ denote the
expectation values and $\theta$ is the angle between {\bf n} and {\bf
n}$^{\prime}$. The distribution
of galaxies in samples 1 -- 4 is binned into pixels exactly the same as
those used for the X-ray maps. Galaxies in sample 5 have also been 
organized into the present pixels, although it required regrouping the
galaxies from the original 10\,arc\,min pixels. Galaxies in the SW pixel
have been redistributed into new ones proportionally to the overlapping
pixel areas. This procedure effectively smoothed out the original data
over a scale comparable to the pixel size. The binned galaxy data
form arrays analogous to the X-ray count rate distributions, where
$\rho_{g}(i)$ is equal to 0, 1, 2,.. according to the number of
galaxies found in the $i$-th pixel. 
To estimate $w_{Xg}(\theta)$ the expectation values in eq.
(1) are substituted by their respective averages obtained from the
data:
\begin{equation}
W_{Xg}(\theta) = {{1 \over n_{ij}(\theta)} \sum_{ij} \rho_X(i)
\rho_{g}(j) \over \overline{\rho_X}\; \overline{\rho_g}} - 1,
\end{equation}
where $\rho_X(i)$ is the count rate in the $i$-th pixel and the sum
extends over all pixel pairs with centers separated ($\theta - 6$\,arc\,min)
and ($\theta + 6$\,arc\,min); $n_{ij}$ is the number of such pairs in
the data and $\overline{\rho_X}$ and $\overline{\rho_g}$ are the average
X-ray count rate and galaxy density, respectively.

\subsection{Numerical results}

\begin{figure}
\psfull
\psfig{file=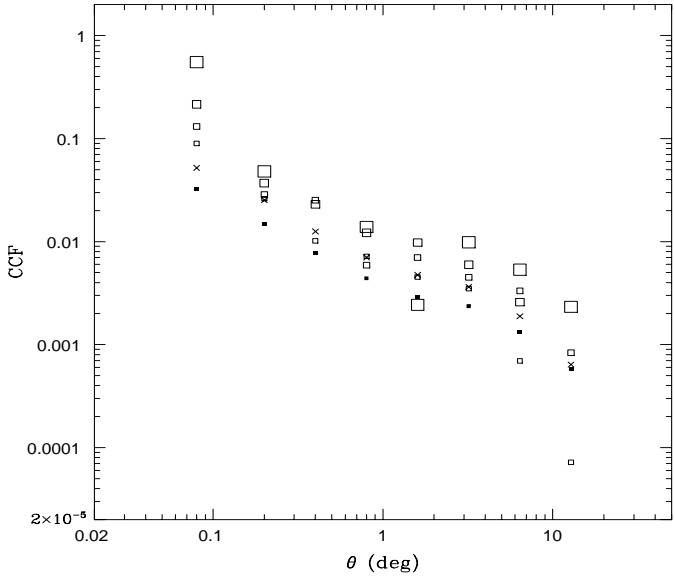,height=7.5cm,width=8.8cm,angle=0}
\caption[ ]{The raw cross-correlation functions (CCFs) of five galaxy samples
and the RASS X-ray maps. Open squares denote four samples from the PGC, filled
squares the SW galaxy counts (sample 5).  The size of the open symbols
corresponds to the apparent magnitude of galaxies: largest squares -- sample 1,
smallest squares -- 4 (see Table 1). The CCFs shown with squares represent the
data without the Abell clusters, crosses show the CCF of the SW and RASS data
including Abell clusters}
\label{Fig1}
\end {figure}

\ni Correlations between the X-ray distribution and all galaxy samples
listed in Table 1 have been computed and the results are shown in
Fig.\ref{Fig1}.
Open squares refer to samples selected from the PGC (samples 1 through 4)
and filled small squares to galaxies from the SW counts. The size of the
symbols corresponds to the apparent magnitude of galaxies in the sample:
the largest squares represent sample 1, the smallest -- sample 4. The CCF
points at the lowest separation give the zero-lag correlations, i.e. $i =
j$ in eq.\,4. All the CCFs shown with squares refer to data which
do not contain the Abell clusters. Pixels close to the position of all Abell
clusters have been removed from the RASS maps and the galaxy samples. The size
of the removed areas has been scaled with the cluster distance class (DC).
For the most distant clusters (DC = 5 and 6), the pixel containing the
cluster and 8 surrounding pixels have been deleted. For DC = 3 and 4 a
radius of 2 pixels was used and for DC = 2 and 1 the areas with radius of 3
and 4 pixels, respectively, were excluded. Crosses in Fig.\ref{Fig1} represent
the CCF of the sample 5 using all the data, i.e. including Abell clusters.
The effect produced by Abell clusters is pronounced over a wide range of
separations: the ratio of the CCF amplitude calculated without clusters to
that using all the data amounts to $\sim 0.6$ and is roughly constant for
$\theta < 10^{\circ}$. Sample 4 is less affected by Abell clusters and
the ratio of ACF amplitudes reaches $0.8$. Galaxies in samples 1, 2 and 3
are virtually not correlated with Abell clusters and the relevant CCFs
do not differ significantly. The strong contribution of Abell clusters to the
galaxy--XRB correlations accentuates the cluster contribution to the 
local luminosity density correlated with the galaxy population. To reduce
the cluster signal in the subsequent analysis we use the data without Abell
clusters. Since the Abell catalogue is not statistically complete and many
smaller galaxy groups are also X-ray emitters, the excision of just the Abell
clusters removes a fraction of X-ray emission produced by intracluster gas
from the RASS maps. In effect, we are unable to eliminate completely
clusters from our analysis and the signal in the respective CCFs represents
the sum of both contributions.

\setcounter{figure}{0}
\renewcommand{\thefigure}{2\alph{figure}}

\begin{figure}
\psfull
\psfig{file=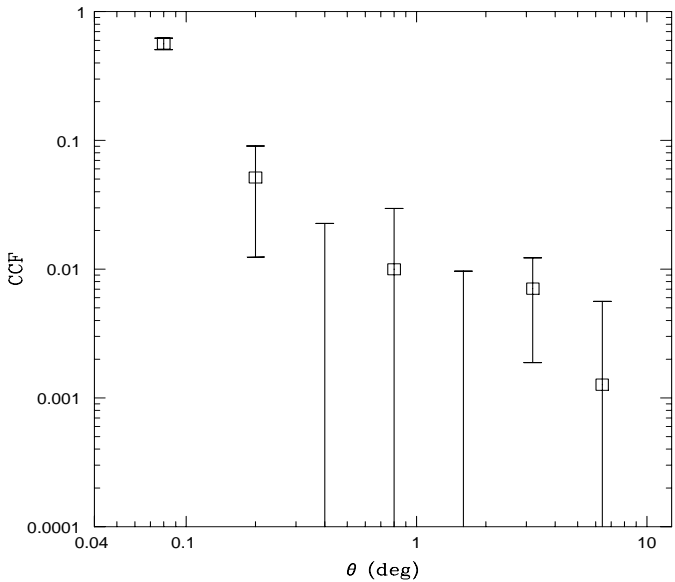,height=7.5cm,width=8.8cm,angle=0}
\caption[ ]{The net CCF (viz. raw CCF -- simulated CCF) of the PGC
sample 1 and the RASS map. Error bars represent rms in the simulations} 
\label{Fig2a}
\end {figure}

\begin{figure}
\psfull
\psfig{file=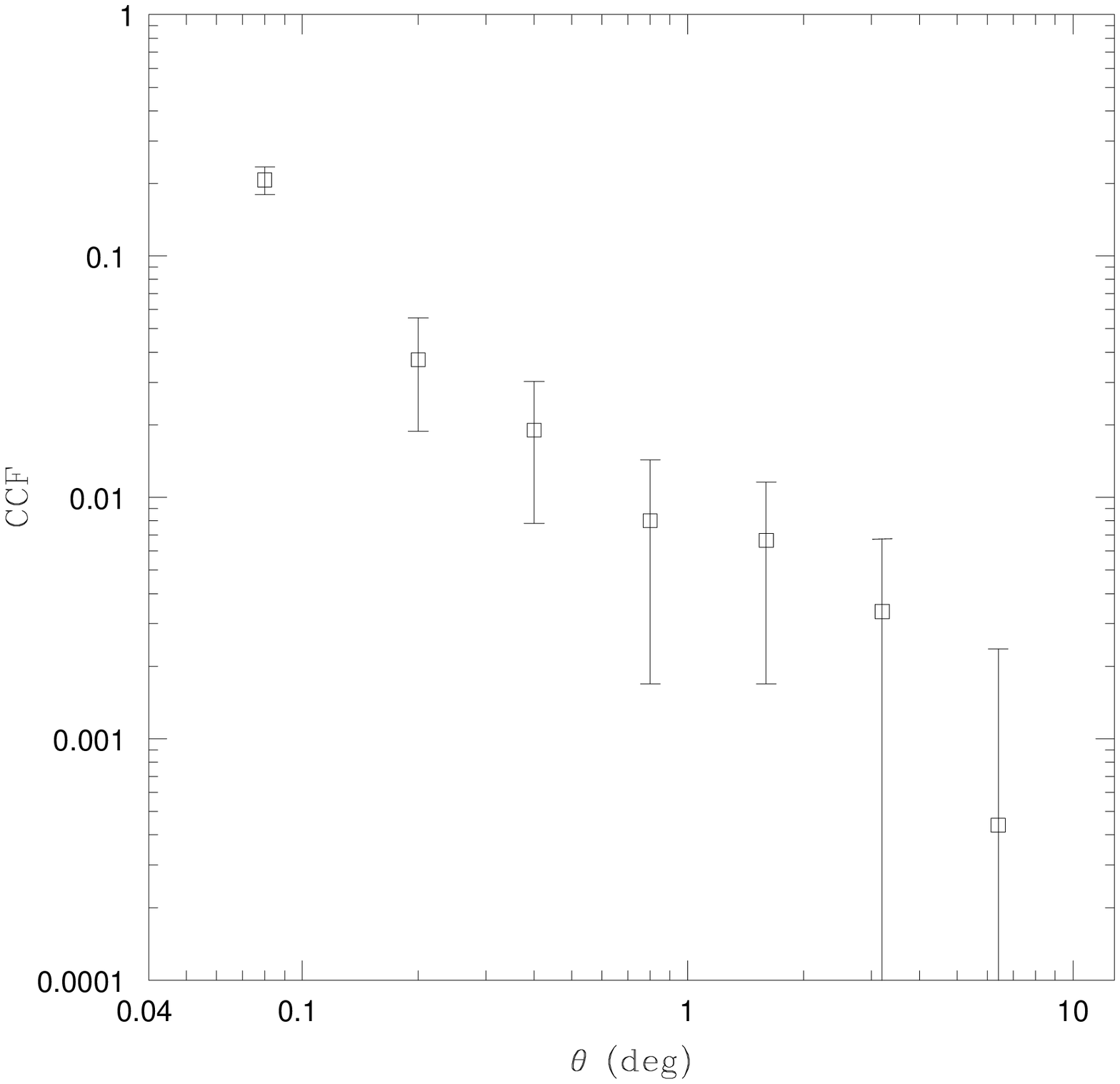,height=7.5cm,width=8.8cm,angle=0}
\caption[ ]{Same as Fig. \ref{Fig2a} for sample 2} 
\label{Fig2b}
\end {figure}

\begin{figure}
\psfull
\psfig{file=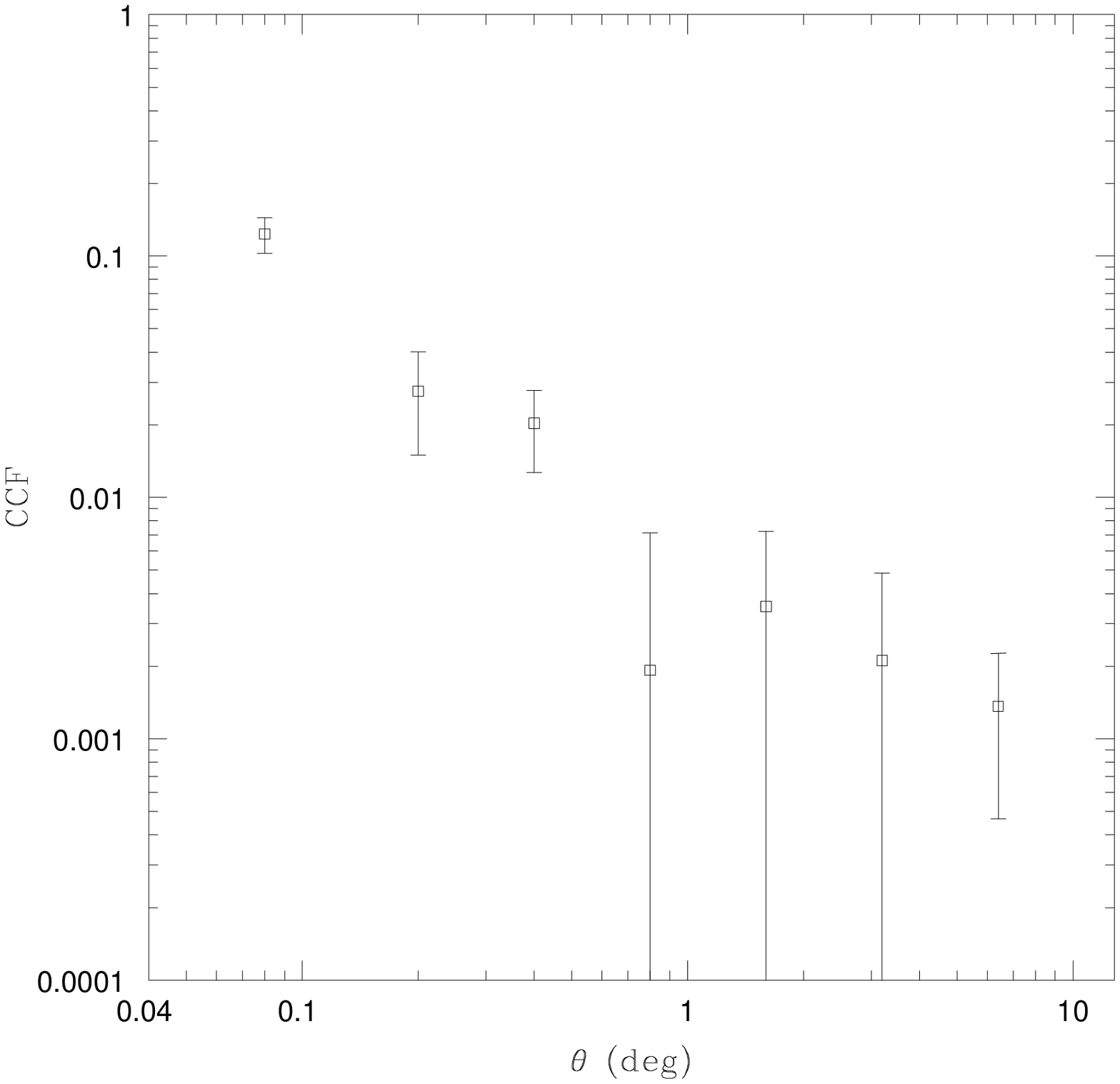,height=7.5cm,width=8.8cm,angle=0}
\caption[ ]{Same as Fig. \ref{Fig2a} for sample 3}
\label{Fig2c}
\end {figure}

\begin{figure}
\psfull
\psfig{file=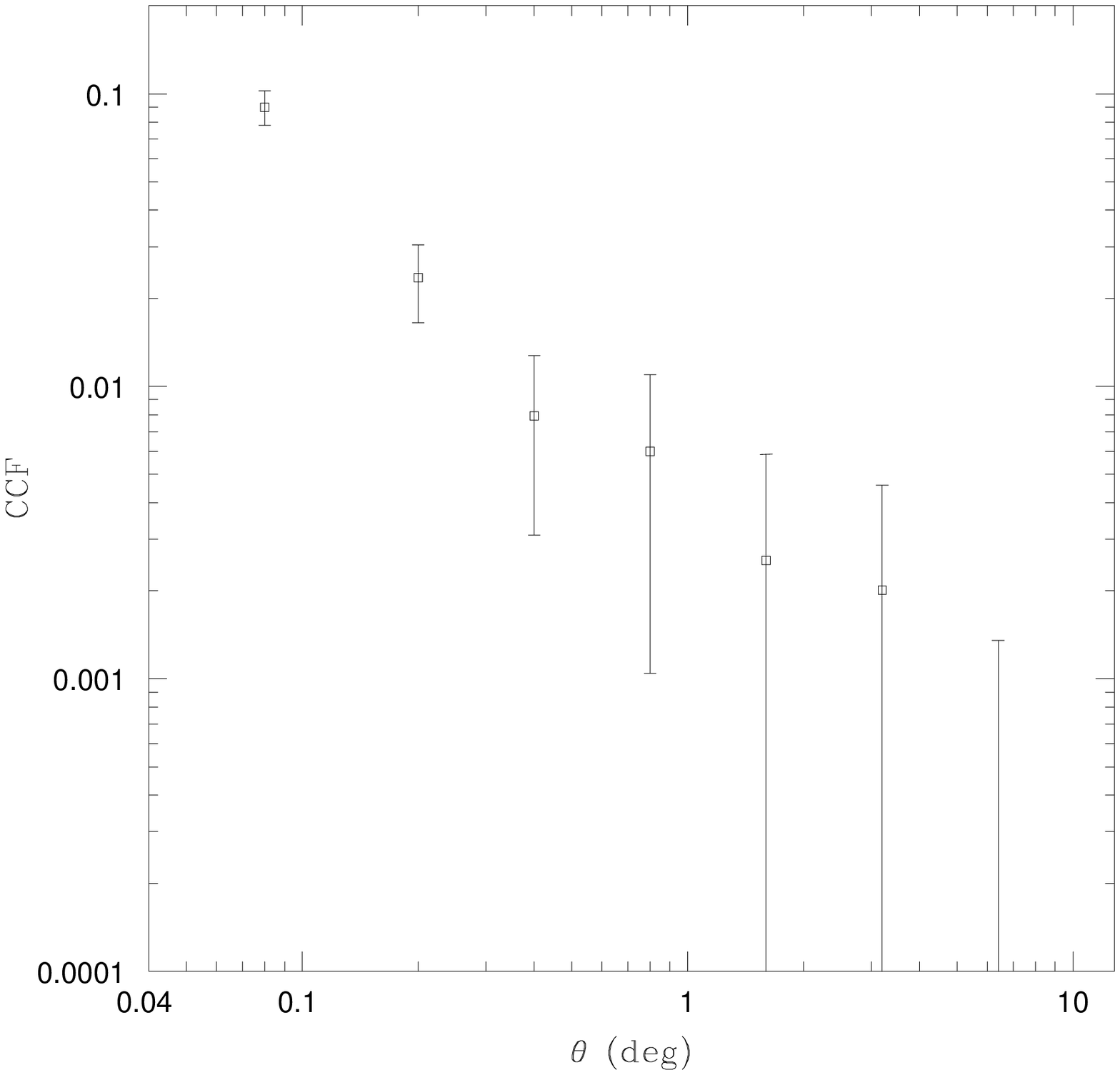,height=7.5cm,width=8.8cm,angle=0}
\caption[ ]{Same as Fig. \ref{Fig2a} for sample 4}
\label{Fig2d}
\end {figure}

\begin{figure}
\psfull
\psfig{file=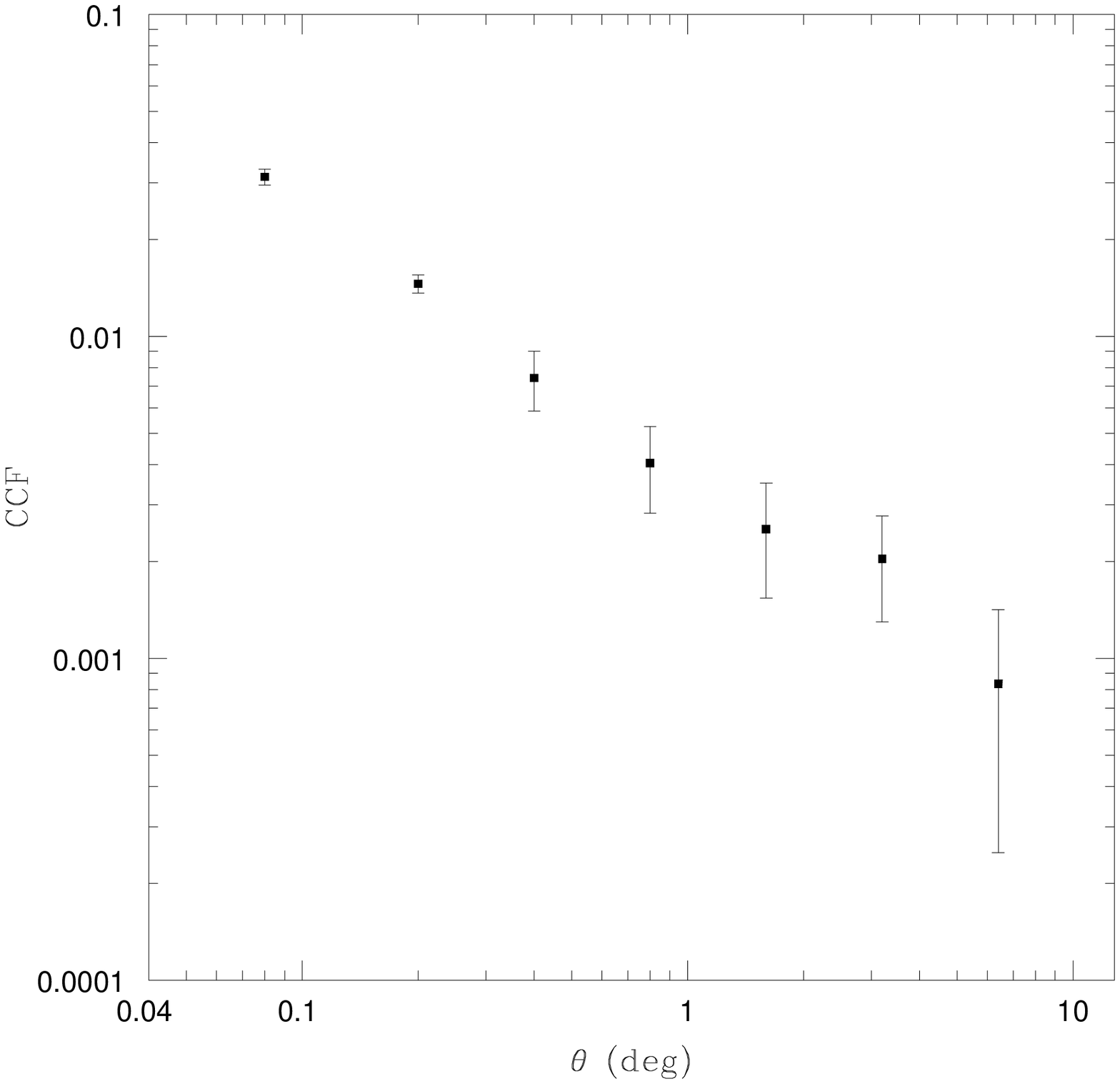,height=7.5cm,width=8.8cm,angle=0}
\caption[ ]{Same as Fig. \ref{Fig2a} for sample 5}
\label{Fig2e}
\end {figure}

Some scatter resulting from statistical fluctuations is visible,
particularly in sample 1 which contains only 108 galaxies.  Despite
this scatter, there are conspicuous systematic trends. The shapes of all CCFs
are similar over a wide range of separations. Below $\sim 1^{\circ}$ the
CCF amplitudes decrease uniformly, while between $1^{\circ}$ and
$10^{\circ}$ all the CCFs exhibit some flattening. At the smallest
separations ($\theta < 0^{\circ}\!\!.3$), where the signal-to-noise
ratio is high, the CCF amplitudes decrease systematically from sample 1
to 5. Substantial errors at larger separations make this trend less apparent
but it is still visible. The uncertainties of our CCF measurements are
difficult to determine. The nonuniform distribution of galaxies on the
celestial sphere combined with relatively high fluctuations of the XRB
on scales up to several degrees generate quite large uncertainties of
the correlation functions. One should point out that actual CCF
uncertainties are significantly larger than those expected from the rms
X-ray count rate scatter and random distribution of galaxies.
We discuss this question in Paper I and find that reasonable
estimates of total errors could be obtained by means of simulations.
Several randomized CCFs are generated using the original X-ray data and
galaxy maps rotated around the galactic polar axis. This method removes
correlations between the data but does not affect the statistical
properties of each distribution. The amplitudes of simulated CCFs are
produced just by random coincidences of fluctuations in both data sets.
It is postulated that the rms scatter between simulations represents total
uncertainties of the CCF estimates. We note, however, that the average
simulated CCFs are systematically positive, while one should expect
symmetric fluctuations around zero. Because a rotation of the
galaxy data around the galactic polar axis does not eliminate a
dependence of the galaxy distribution on the galactic latitude,
positive amplitudes of simulated CCFs suggest that there is also some
systematic trend of the X-ray data with galactic latitude. To correct
for this effect, we have subtracted from the observed correlations the
average simulated amplitudes in each sample. At small separations the
extragalactic signal dominates and the correction is small, but above
$\sim 1^{\circ}$ the galactic contamination is non-negligible. Results
for 5 samples are shown in Figs.\,2\,a--e (where the symbols refer to the 
same samples as in Fig. \ref{Fig1}). Error bars correspond to the rms
scatter between simulations. In sample 5 the net correlation extends
clearly to several degrees. For nearby samples with relatively small
number of galaxies uncertainties are comparable to the signal on scales
above $\sim 1^{\circ}$. 

\section{Galaxy distribution vs. the X-ray background fluctuations}

\ni The CCFs obtained in the previous section describe the coupling
between the galaxy distribution and the intensity of the X-ray
background radiation. The amplitude $W_{Xg}(\theta)$ gives the relative
enhancement of the X-ray intensity at separation $\theta$ of a randomly
chosen galaxy. At small separations, particularly for zero-lag, this
enhancement is produced mostly by the individual galaxy itself, while at larger
distances the signal results from other X-ray sources which are
correlated with galaxies from the sample.

The average distribution of galaxies around a randomly chosen object is
described by galaxy correlation functions. Thus, the CCFs between the
X-ray data and the galaxy distribution shown on Figs.\,2a--e depend on
the galaxy--galaxy correlation functions. The formulae of the
auto-correlation function of galaxies in a single sample and the
cross-correlation function between two galaxy samples are analogous to
those given by eqs.\,(3) and (4): the ACF of galaxies in the sample $k$
($k = 1,...,5$), $w_{kk}(\theta)$ is defined as follows:
\begin{equation}
w_{kk}(\theta) = {\langle \rho_k(\mbox {\bf n}) \rho_k
(\mbox {\bf n}^{\prime}) \rangle_{\theta} \over
\langle \rho_k \rangle^2} - 1,
\end{equation}
and the CCF between samples $k$ and $l$, $w_{kl}(\theta)$:
\begin{equation}
w_{kl}(\theta) = {\langle \rho_k(\mbox {\bf n}) \rho_l
(\mbox {\bf n}^{\prime}) \rangle_{\theta} \over
\langle \rho_k \rangle \langle \rho_l\rangle} - 1.
\end{equation}

The number of excess galaxies per pixel above the average concentration
from sample $l$ at separation $\theta$ from a randomly chosen galaxy in
sample $k$ is described by the appropriate CCF (e.g. Peebles 1980):
\begin{equation}
n_{kl} = \langle \rho_l\rangle \cdot w_{kl}(\theta).
\end{equation}
Let $f_l^{opt}$ denote the average optical flux produced by a galaxy
from the sample $l$. Then the total enhancement of the optical surface
brightness of the sky around a randomly chosen galaxy in the sample $k$,
$\Delta f_k^{opt}$, due to galaxies in all the samples is given by:
\begin{equation}
\Delta f_k^{opt}(\theta) = A_k^{opt}(\theta) + \sum_{l} f_l^{opt}
\langle \rho_l\rangle w_{kl}(\theta),
\end{equation}
where term $A_k^{opt}(\theta)$ accounts for the finite size of galaxies
and denotes the average optical flux produced by the chosen galaxy in a
pixel at a separation $\theta$ and the sum extends over all galaxy
samples which exhibit correlations with the sample $k$.

The CCF between galaxies in sample $k$ and the X-ray distribution
gives the average excess of the X-ray intensity around a randomly chosen galaxy:
\begin{equation}
\Delta f_k^{X}(\theta) = \langle \rho_X \rangle w_{Xk}(\theta).
\end{equation}
\ni The ratio of fluctuation amplitudes in both domains (X-ray and optical)
defined by eqs. (8) and (9) is used to determine the galaxy contribution
to the XRB in the next section.

\subsection{Approximate solution}

\ni We define the optical flux by means of B magnitudes as:
\begin{equation}
f^{opt} = \nu_B\,f_{\nu_B}
\end{equation}
where $\nu_B = 6.8\times 10^{14}$\,Hz is the effective frequency of
the B band\footnote[1]{Note that flux in the optical domain in various
papers is defined differently, e.g. Maccacaro et al. (1988) define the
optical flux as: $\log f_v = - 0.4 m_{\rm v} - 5.37$ while our eq.\,(10)
is equivalent to $\log f^{opt} = - 0.4 m_{\rm B} - 4.57$.}.
We have used the following conversion from the B band to
energy flux $f_{\nu_B}$ in erg\,s$^{-1}$\,cm$^{-2}$\,Hz$^{-1}$: $\log
f_B = - 19.41 - 0.4$\,B (Butcher et al. 1980). Note, that the zero point
of this normalization as well as the particular definition of
$f^{opt}$ in eq.\,10 do not affect our subsequent estimates of the
galaxy contribution to the XRB. This is because in the
calculations we utilize effectively the ratio of X-ray to optical fluxes
and the ratio of corresponding absolute luminosities. It requires, however,
consistent definitions of optical apparent flux and absolute
luminosity in the B band (see eq.\,12).

Because our galaxy data collected in five samples cover an apparent magnitude
range between 10 and $\sim 18.6$, eq.\,(8) for each $k = 1, ..., 5$
describes only the fraction of optical fluctuations which is associated
with galaxies in these samples. The bright magnitude constraint
of 10 is not restrictive because the small number of galaxies with 
$m < 10$ does not
affect the observed correlations. The problem of galaxies not included
in the calculations is severe at the faint end. Numerous galaxies below
the SW count threshold influence the observed CCFs. The effect is
strongest for sample 5, but could also be
important for other samples. In the first approximation we ignore all the
cross-correlation terms in eq.\,(8) and use only the ACFs. We avoid
problems with the point response function and calculations of $A_k$ by 
using the CCFs integrated over separations $\theta < 18^{\prime}$,
a distance significantly larger than the effective angular resolution of
the RASS. In this case, $A_k(\theta)$ is replaced by $f_k^{opt}$ and
eqs.\,(8) can be rewritten in the form:
\begin{equation}
\Delta f_k^{opt}(\theta) = \sum_{l} f_l^{opt} (\langle \rho_l\rangle 
W_{kl} +\delta_{kl}),
\end{equation}
where $W_{kl}$ denotes the CCF estimator defined by eq.\,(4),
$\delta_{kl} = 1$ for $l = k$ and $\delta_{kl} = 0$ otherwise.
Amplitudes $W_{kl}$ are calculated using overlapping
pixels (zero lag) and the nearest neighbours (first two points on
Figs.\,2a-e).

The average count rate defined by eq.\,(2) in the discussed area $\langle
\rho_X \rangle = 74.6\times 10^{-6}$\,PSPC cnt\,s$^{-1}$\,arcmin$^{-2}$.
Conversion rates of PSPC counts to the flux units are given 
for various spectra in Paper I. In the following calculations we adopt a
conversion factor corresponding to a power law spectrum with an
energy spectral index of $-1$ and a hydrogen column density of
$1.8\times 10^{20}$\,cm$^{-2}$, as these parameters are representative
for our energy band and sky region (Hasinger 1992, Paper I). Using the standard
calibration of the PSPC we find that 1\,PSPC cnt\,s$^{-1}$ 
is equivalent to a flux outside the Galaxy of $3.29\times
10^{-11}$\,erg\,cm$^{-2}$\,s$^{-1}$ in the 0.5 - 2.0\,keV energy band.

\begin{table*}
\caption[ ]{X-ray and optical fluxes (approximate solution)}
\vspace*{0.10cm}
\begin{tabular}{c|c|c|l|c}
\hline
Sample& $\Delta f_k^X$      & $\Delta f_k^{opt}$(ACF)&
                                 $\Delta f_k^X / \Delta f_k^{opt}$(ACF) &
                                                                      S/N \\
 $k$& \multicolumn{2}{|c|}{[erg\,s$^{-1}$\,cm$^{-2}$]}&          &        \\
  1 & $3.45\times 10^{-13}$ & $8.74\times 10^{-10}$ & $0.00040$  & 1.23   \\
  2 & $1.78\times 10^{-13}$ & $1.61\times 10^{-10}$ & $0.0011$   & 2.02   \\
  3 & $1.22\times 10^{-13}$ & $4.92\times 10^{-11}$ & $0.0025$   & 2.20   \\
  4 & $9.83\times 10^{-14}$ & $2.61\times 10^{-11}$ & $0.0038$   & 3.36   \\
  5 & $5.23\times 10^{-14}$ & $1.05\times 10^{-11}$ & $0.0050$   & 15.5   \\
\hline
\end{tabular}
\end{table*}

The fluctuation amplitudes in the X-ray band obtained using eq.\,(9) with the
$W_{Xk}$ estimators defined by eq.\,(4) are given in column 2 of Table 2.
Fluctuations in the optical domain produced by galaxies in the sample
under consideration are calculated by means of eq.\,(11) using only the
auto-correlation term i.e. $k=l$ are denoted by $\Delta f_k^{opt}$(ACF) and
are given in column 3. The optical data quoted in Table 2 ignore the
nonuniform distribution of galaxies in other samples and
underestimate the actual optical fluctuations. In column 4 the ratio
of X-ray to B-band fluctuations
$R_k^{ACF} = \Delta f_k^X/\Delta f_k^{opt}$(ACF)
is listed. Systematic variations of $R_k^{ACF}$ by a factor of $\sim 12$
indicate that the CCF terms in eq.\,(11) are significant. A signal-to-noise
ratio of the galaxy--X-ray CCF determinations used to compute
$\Delta f_k^X$ is listed in column 5.

\subsection{Model galaxy distribution}

\ni The distributions of galaxies in the different magnitude ranges are highly
correlated. This is because the galaxy luminosity function extends over
several magnitudes and galaxies occupying the same volume of space can be
members of various magnitude limited samples. Also, the apparent magnitudes
of galaxies listed in the PGC are subject to fairly large errors which
additionally spread spatially intermingled galaxies over different
samples. Cross-correlations between samples increase the total fluctuation 
amplitude of the optical light substantially. Using the CCFs between
our 5 galaxy samples we can directly measure fluctuations associated with
galaxies brighter than B $\approx 18.6$. Effects produced by fainter
galaxies are estimated using the model distribution of galaxies.

Estimates of the angular correlation amplitudes of galaxies in the
magnitude limited samples are obtained assuming standard 3D galaxy
correlation function in the form: $\xi(r) = (r/r_{\circ})^{-\gamma}$.
It has been assumed that $\xi(r)$ is independent of the absolute
luminosity of galaxies. We have used flat cosmological model with
$H_{\circ} = 100$\,km\,s$^{-1}$Mpc$^{-1}$, $\Lambda = 0$ and $q_{\circ}
= 0.5$. To generate the redshift distribution of galaxies at given
apparent magnitude and subsequently, to calculate the galaxy angular
correlation functions, the galaxy luminosity function (LF) is needed.
In the calculations the LF has been approximated by the Schechter
(1976) function with $\rm M_{\rm B}^{\ast} = -19.49$ and $\alpha =
-1.5$. Driver et al. (1994) discussed various models of the galaxy
luminosity distributions consistent with the galaxy counts. Parameters
used in the present paper are adopted from the Driver et al. model
based on a single Schechter function. We have also used their approximations
for $K$-corrections. Even with such simplified model we have been able
to reproduce fairly well the ACF amplitudes measured in our galaxy samples.
To fit the predicted amplitudes to the observed ones we varied only
the spatial correlation length $r_{\circ}$ and
the best agreement has been found for $r_{\circ} = 4.0$\,Mpc.
In Table 3 we compare the observed and calculated ACF amplitudes
averaged over 9 pixels.

\begin{table*}
\caption[ ]{The ACF amplitudes averaged over $36^{\prime}\times 36^{\prime}$}
\vspace*{0.10cm}
\begin{tabular}{c|c|c}
\hline
Sample& $W_{kk}$ - observed & $w_{kk}$ - model \\
  1   &     5.76            &   5.55           \\
  2   &     3.79            &   3.66           \\
  3   &     1.48            &   1.96           \\
  4   &     1.04            &   1.06           \\
  5   &     0.23            &   0.26           \\
\hline
\end{tabular}
\end{table*}

Using the present model we have calculated the CCFs between the galaxy
samples 1 -- 5 and galaxies fainter than $\rm m_{\rm B} = 18.6$.
In the computations we have divided the latter galaxies into 5
magnitude bins with $\Delta \rm m_{\rm B} = 1$:
18.6 -- 19.6 -- ... -- 22.6 -- 23.6. Substituting the actual CCF
amplitudes $W_{kl}$ for $k, l = 1,..., 5$ and model amplitudes for $k =
1,..., 5$ and $l = 6, ..., 10$ into eq.\,(11) we calculated the average
fluctuations of the sky brightness in the B band around sample galaxies
down to ${\rm m}_{\rm B} = 23.6$. Effects produced by fainter galaxies
were estimated using extrapolation from ${\rm m}_{\rm B} < 23.6$.
Contributions of ${\rm m}_{\rm B} > 23.6$ galaxies relative to those
brighter than 23.6 are below 1\,\% for fluctuations around galaxies in
three brightest galaxy samples (k = 1, 2, 3) and reach 1.6\,\% and
4.9\,\% for samples 4 and 5, respectively.  Results of these
calculations are given in Table 4.

\begin{table*}
\caption[ ]{Galaxy contribution to the XRB}
\vspace*{0.10cm}
\begin{tabular}{c|c|c|l|l}
\hline
Sample& $\Delta f_k^{opt}$(m$_{\rm B} < 18.6$) &
                  $\Delta f_k^{opt}$(total)      &
                            $\Delta f_k^X /\Delta f_k^{opt}$(total) &
                                                                $ C_k$\\
 $k$ & \multicolumn{2}{|c|}{[erg\,s$^{-1}$\,cm$^{-2}$]} &     &       \\
  1  & $9.22\times 10^{-10}$ & $9.25\times 10^{-10}$&$0.00037$& 0.071 \\
  2  & $2.18\times 10^{-11}$ & $2.26\times 10^{-10}$&$0.00079$& 0.15  \\
  3  & $8.69\times 10^{-11}$ & $9.53\times 10^{-11}$&$0.0013$ & 0.24  \\
  4  & $4.81\times 10^{-11}$ & $5.67\times 10^{-11}$&$0.0017$ & 0.33  \\
  5  & $1.67\times 10^{-11}$ & $2.59\times 10^{-11}$&$0.0020$ & 0.38  \\
\hline
\end{tabular}
\end{table*}

Observed average amplitudes of optical fluctuations produced by galaxies
brighter than $\rm m_{\rm B} = 18.6$ are listed in column 2.
Estimates of total fluctuations $\Delta f_k^{opt}$(total)
and X-ray--to--optical ratios
$R_k^{tot} = \Delta f_k^X /\Delta f_k^{opt}$(total)
are given in columns 3 and 4, respectively. We note that $R_k^{tot}$
exhibits still large variations, although it is substantially more
stable than $R_k^{ACF}$. We discuss variations of $R_k^{tot}$ in the
next section.

The ratio of apparent fluctuations $R_k^{tot}$ is equal to the ratio of
X-ray--to--optical volume emissivities ${\cal L}_X/{\cal L}_{opt}$.
We calculate the optical volume emissivity in the B-band integrating
the LF over all the optical luminosities:
\begin{eqnarray}
\lefteqn {
{\cal L}_{opt} = \nu_{\rm B} \int\!L_{\nu_{\rm B}}\,\varphi(L_{\nu_{\rm B}})
\,dL_{\nu_{\rm B}} = n^{\ast}\,\nu_{\rm B}\,L_{\nu_{\rm B}}^{\ast}
\,\Gamma(\alpha + 2)} \nonumber \\
 & &  = 5.20\times 10^{41}\,{\rm erg\,s}^{-1}\,{\rm Mpc}^{-3},
\end{eqnarray}
where $n^{\ast} = 1.47\times 10^{-2}$\,Mpc$^{-3}$ is the normalization
of the Schechter LF, $L_{\nu_{\rm B}}^{\ast} = 2.92\times 10^{28}$
\,erg\,s$^{-1}$Hz$^{-1}$ corresponds to ${\rm M}_{\rm B}^{\ast} = - 19.49$.
The galaxy counts using these parameters are in good agreement with the
actual number of galaxies in our samples. We now calculate the X-ray
volume emissivity ${\cal L}_X(gal)_k$ correlated with the galaxy
distribution for each sample:
\begin{equation}
{\cal L}_X(gal)_k = R_k^{tot}\, {\cal L}_{opt}.
\end{equation}
To assess contribution of this local emission to the total XRB, we note
that without cosmological evolution 
${\cal L}_X(total) = 2.73\times 10^{39}$\,erg\,s$^{-1}$\,Mpc$^{-3}$
in the 0.5 -- 2.0\,keV band is required to produce the observed intensity
of the XRB (Hasinger et al. 1993, Paper I). In column 5 of
Table 4 we list the ratio:
\begin{equation}
C_k = {\cal L}_X(gal)_k / {\cal L}_X(total),
\end{equation}
Thus, $C_k$ denotes the fractional contribution to the XRB by
a nonevolving population of sources generating a constant luminosity
${\cal L}_X(gal)_k$ per unit volume of co-moving space
integrated to large redshifts (e.g. Paper I, eq.\,18).

Estimates of the local volume X-ray emissivity and its contribution to the
XRB based on 5 galaxy samples still cover an uncomfortably wide range. Albeit
this is distinctly smaller than variations of $R_k^{ACF}$ found when the
cross-correlation terms were ignored (column 4 in Table 2), present
$C_k$ measurements span from 7\,\% to 38\,\%. These changes are
inconsistent with one value, expected if X-ray properties of each sample
are representative to the whole galaxy population. Growth of $C_k$ with $k$
taken literally would indicate that samples have substantially different
average X-ray luminosities associated with a single galaxy. Obviously, such
interpretation is not accepted and an alternative systematic effect is
proposed in the next section to explain sample-to-sample variations.

\section{Effects of extended emission}

\ni The ratios of X-ray--to--optical fluctuations have been calculated in the
preceding section for separations $\theta < 18^{\prime}$. We now
apply these quantities to synthetize predicted CCF between the
galaxy distribution and the X-ray sky in the wider range of separations and
compare them with the actual measurements. The procedure is as follows.
The amplitude of the optical fluctuations $\Delta f_k^{opt}(\theta)$ is
calculated by means of eq.\,(11) using galaxy ACFs and CCFs
$W_{kl}(\theta)$.  The ``$\delta_{kl}$'' term at the right-hand side of the
equation is used only for zero lag correlations ($i = j$ in eq.\,(4)).
Then the values $\Delta f_k^X(\theta)$ are obtained using the $R_k^{tot}$
from column 4 of Table 4. Finally, predicted CCF $w_{Xk}(\theta)$
for $k = 1,..., 5$ are calculated from eq.\,(9) and the results are
shown in Figs.\,3a-e. Open squares and error bars show the actual
galaxy--X-ray CCF (same as in Figs.\,2a-e), while crosses represent
calculated CCFs according to above prescription. For comparison,
the ACFs in each galaxy sample are shown with filled symbols.

\setcounter{figure}{0}
\renewcommand{\thefigure}{3\alph{figure}}

\begin{figure}
\psfull
\psfig{file=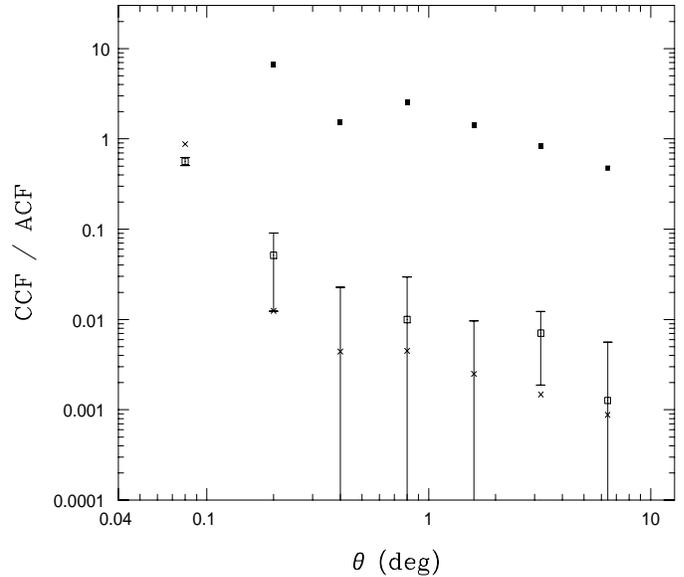,height=7.5cm,width=8.8cm,angle=0}
\caption[ ]{Correlation functions for galaxies in sample 1:
filled squares -- the ACF, open squares and error bars -- the CCF of
sample 1 and the RASS maps (same as Fig. \ref{Fig2a}), crosses -- the CCF
predicted using correlations between all galaxy samples (see text for details)}
\label{Fig3a}
\end {figure}

\begin{figure}
\psfull
\psfig{file=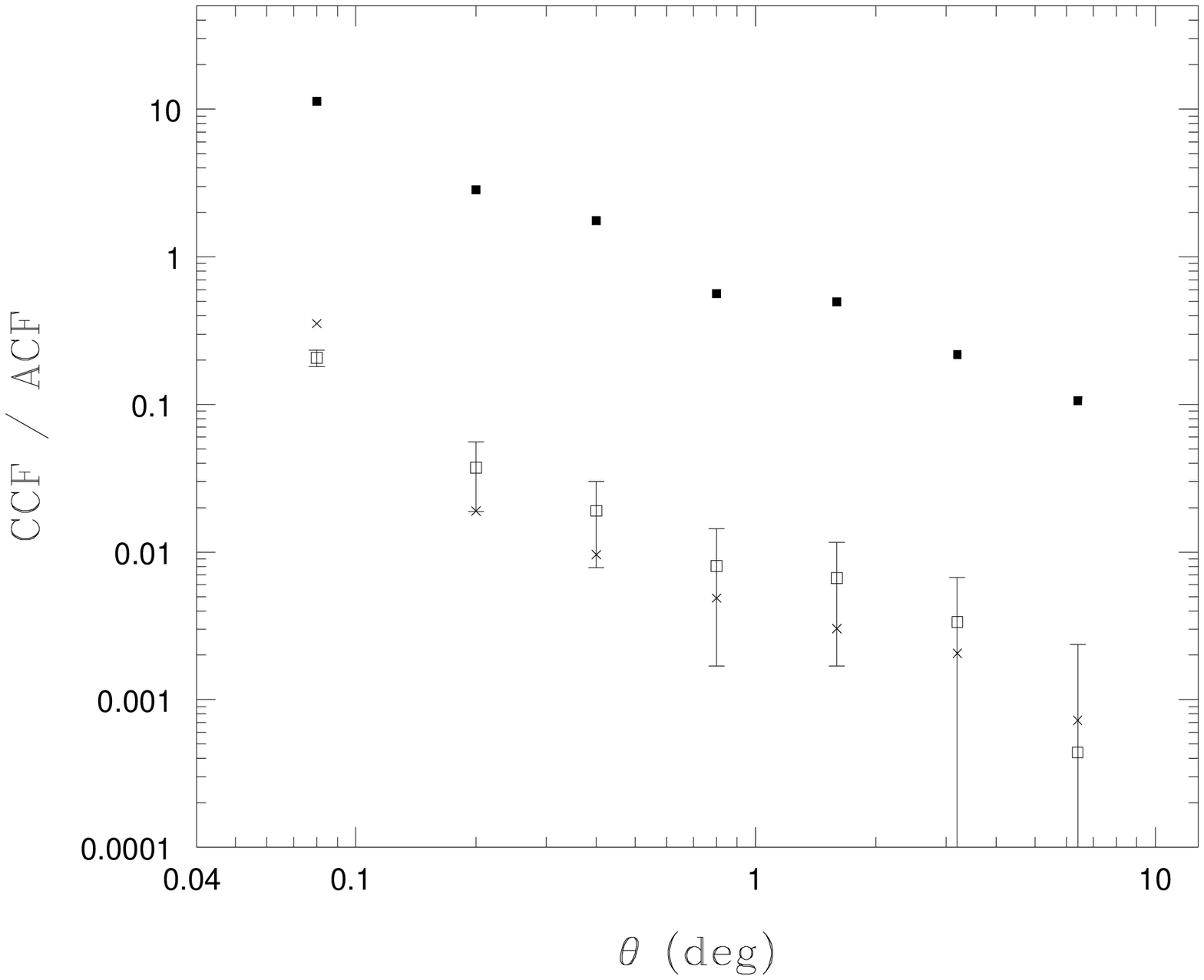,height=7.5cm,width=8.8cm,angle=0}
\caption[ ]{Same as Fig. \ref{Fig3a} for sample 2} 
\label{Fig3b}
\end {figure}

\begin{figure}
\psfull
\psfig{file=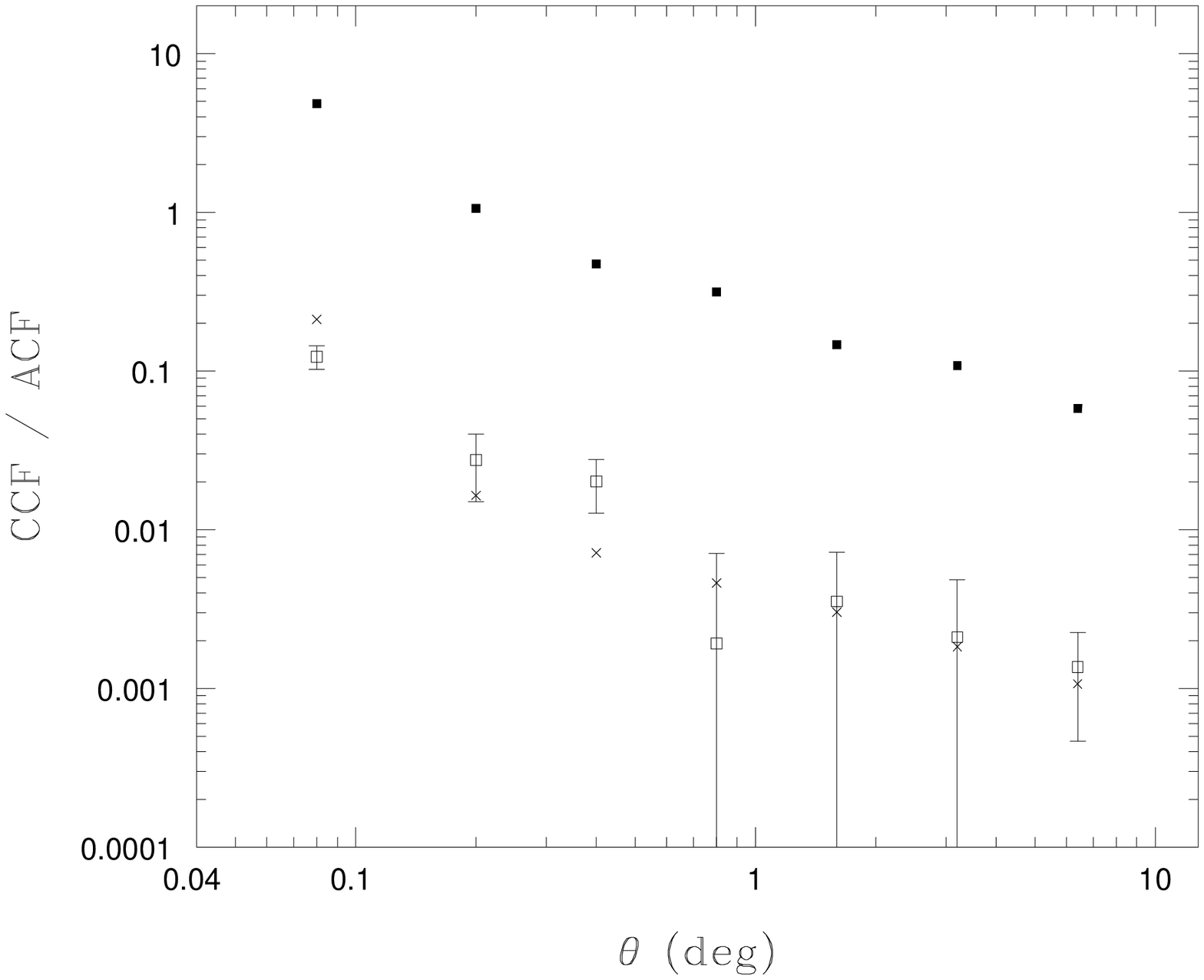,height=7.5cm,width=8.8cm,angle=0}
\caption[ ]{Same as Fig. \ref{Fig3a} for sample 3}
\label{Fig3c}
\end {figure}

\begin{figure}
\psfull
\psfig{file=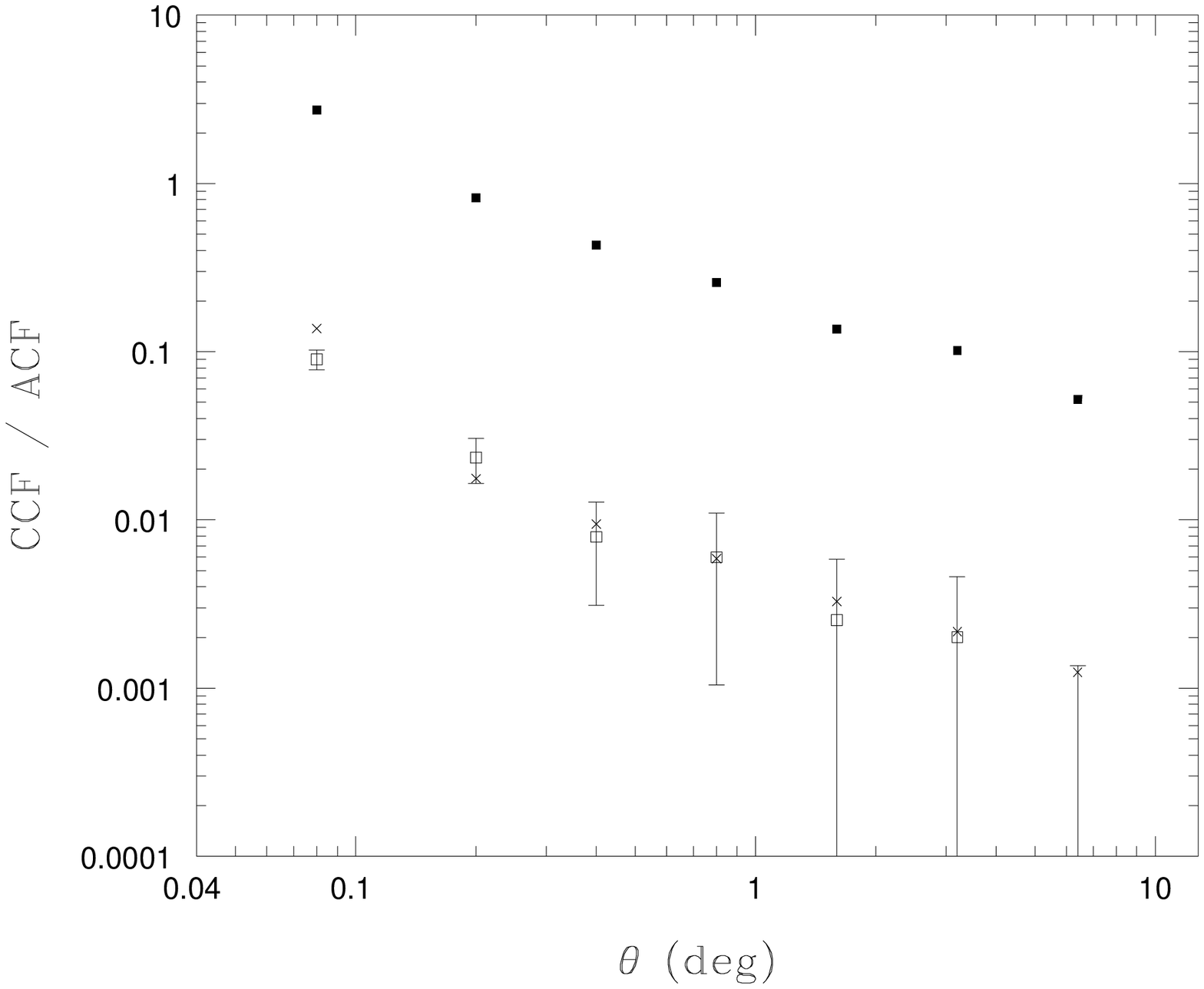,height=7.5cm,width=8.8cm,angle=0}
\caption[ ]{Same as Fig. \ref{Fig3a} for sample 4}
\label{Fig3d}
\end {figure}

\begin{figure}
\psfull
\psfig{file=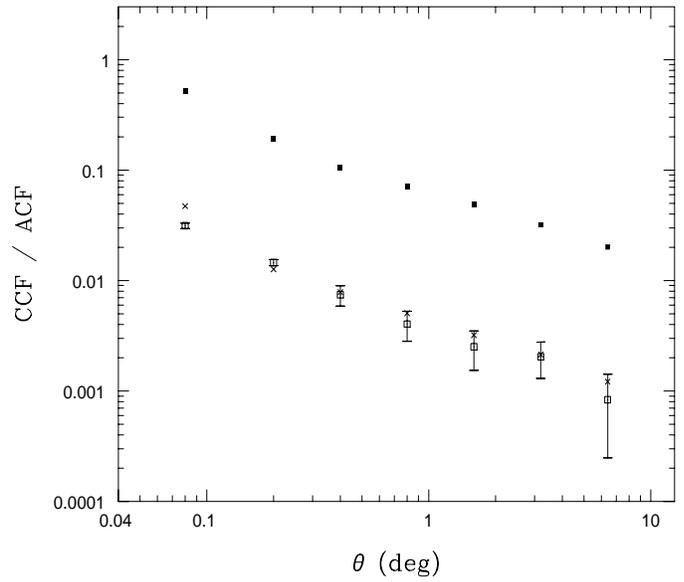,height=7.5cm,width=8.8cm,angle=0}
\caption[ ]{Same as Fig. \ref{Fig3a} for sample 5}
\label{Fig3e}
\end {figure}

Results for sample 1 (Fig.\ref{Fig2a}) do not allow to make
conclusive statements due to excessive uncertainties. We note only that
observed and synthetized ACFs in sample 1 are in qualitative agreement and
the subsequent discussion is limited to the samples $k = 2 - 5$.
At separations below $18^{\prime}$ all calculated CCFs exhibit similar
deviations from the actual measurements: the synthetized zero-lag point
is situated above the real one, while the relative positions of
the next point ($6^{\prime} < \theta < 18^{\prime}$) are reversed.
This is because the predicted galaxy--X-ray CCFs are obtained
under the assumption that the galaxy residing in the zero-lag pixel
does not produce X-ray signal in the surrounding pixels. This condition
is not satisfied in the real CCF due to the finite width of the point
spread function.

Comparison of differences between the simulated and observed CCFs for
samples $k = 2 - 5$ offers a possible explanation of systematic changes
of the X-ray--to--optical ratio. The predicted CCF amplitude in sample
3 at $\theta < 0.\!\!^{\circ}5$ is below the observed CCF. A similar
effect is present in the sample 2 also at larger separations, although
within the error bars. Samples 4 and 5 show near perfect agreement
between observations and predictions. These data indicate,
albeit at a relatively low significance level, that the total
angular extent of X-ray emission associated with galaxies has been
underestimated in our calculations in samples 2 and 3, while galaxies
in samples 4 and 5 are more distant and their sizes are below the pixel
size. To check quantitatively this effect, we have recalculated the
X-ray--to--optical ratio and the relative galaxy contribution $C_k$ to the
XRB using the first three points instead of two of the ACFs and CCFs, e.g. the
correlation functions have been averaged over $\theta < 30^{\prime}$ rather
than $18^{\prime}$ and the calculations described in Section 5 have been
repeated. One could expect that this change of the angular separations used
in the analysis will not affect results in the samples 4 and 5. In fact, we
get now $C_k = 0.31$ and $0.37$, respectively. Differences with our
previous estimates of $0.33$ and $0.38$ (Table 4) are clearly within the
uncertainties. In the sample 3 we obtain substantially increased X-ray
contribution, namely $C_3 = 0.35$, as compared to $0.24$ for $\theta <
18^{\prime}$. A similar but weaker effect is detected in the sample 2:
$C_2(<30^{\prime}) = 0.18$, while $C_2(<18^{\prime}) = 0.15$. Further
increase of separations does not provide more accurate estimates because of
the large CCF error bars above $\theta \simeq 1^{\circ}$. We conclude that
variations of $C_k$ could be eliminated assuming that the average X-ray
emission by galaxies in sample 3 extends up to $\sim 0.\!\!^{\circ}5$ and
even to larger distances in the sample 2.
One should note that the accuracy of the CCF measurements is quite low
even at these small separations. Inclusion of the $18^{\prime} <
\theta <  30^{\prime}$ point into the $C_k$ estimates does not reduce
the statistical uncertainties involved in our calculations. It shows only
that the data are consistent with the supposition that a substantial
fraction of the flux correlated with galaxies originates in extended
sources. At the same time, the extended emission can explain the
variations of $C_k$ found for $\theta < 18^{\prime}$ in a natural way.

Uncertainties of the relative contribution to the XRB, $C_k$ are produced by
various sources. Both statistical as well as systematic effects are
present. Our error estimates of the galaxy--X-ray CCFs shown in the figures
provide some insight into the problem. These errors are given as
signal--to--noise ratio of the CCF measurements in column 5 of Table 2.
Because they are directly related to the number of galaxies involved in
the computations, sample 5 exhibits the smallest statistical uncertainties.
Two other samples which were effectively used to estimate $C_k$, viz.
samples 3 and 4, suffer from substantially larger uncertainties.
Nevertheless, all three samples eventually provided very similar
estimates of the total galaxy contribution to the XRB. This conspicuous
agreement between three samples shows that our approach constitutes an
efficient and coherent method of calculations. We note that three samples of
galaxies under consideration cover almost 5 magnitudes in the apparent
brightness. The range of redshifts is also very wide. The estimated
characteristic redshift of the most distant galaxies in the sample 5
exceeds $0.1$, while the mean redshift in the sample 3 $\langle  z_3
\rangle = 0.018$. The data are quite heterogeneous: sample 3 and 4 come
from the PGC while sample 5 is a subset of the SW counts. All these
factors as well as the multi--step procedure of the $C_k$ determination make
literal assessment of the final errors unreliable. Due to this reason we
refrain from formal estimates of the $C_k$ uncertainties. However, the results
based on samples 3, 4 and 5 indicate explicitly that the total contribution of
X-ray emission associated with the local galaxy population assuming no 
evolution falls in the range $0.3 \la C_k \la 0.4$. 

The large pixel size does not allow for a detailed study of the spatial
distribution of the emission. Crude estimates of the magnitudes involved
are obtained as follows. Mean galaxy redshifts in samples 2 and 3 are
$0.010$ and $0.018$, respectively. Using the $C_k$ values from Table 4
we find out that roughly half of the X-ray emission by galaxies
in sample 2 comes from separations greater than $\sim 18^{\prime}$
while all the emission in sample 3 comes within $\sim 30^{\prime}$.
Linear sizes corresponding to $\sim 18^{\prime}$ at $z = 0.010$ and
to $\sim 30^{\prime}$ at $0.018$ are $0.16$\,Mpc and $0.47$\,Mpc.
Thus, half of the emission comes from regions of radius $\sim 0.5$\,Mpc
but well outside the optical extent of galaxies. 
We would like to stress once more that the exact nature
of the extended emission introduced to obtain consistent results on the
X-ray luminosity density within our data is not well constrained
by the correlation
analysis. Our estimate that about half of the emission could come from
regions of $\sim 1$\,Mpc size does not imply that each galaxy
is surrounded by such emitting region. The extended signal could be generated
either by weak sources associated with a large number of individual galaxies
or by a small number of stronger sources related to galaxy groups or clusters.
Below we consider such two models and confront them with observational
constraints.

\subsection{Cluster emission}

\ni Extended X-ray emission by small clusters and groups of galaxies
contained in the galaxy samples used in the present investigation
offers a natural explanation for the $\sim 1$\,Mpc size X-ray sources
reported above. We note that only Abell clusters have been
excluded from the data, while all variety of smaller groups as well as some 
unspecified
number of rich clusters missed by Abell are still present in our material.
Obviously, only a fraction of galaxies belongs to those groups or
clusters, but our analysis is unable to distinguish between a large number of
weak extended haloes around every galaxy and less numerous but stronger
X-ray cluster sources. The possible significance of poor cluster emission for
the XRB fluctuations is discussed in some details in Paper I. Using
analogous arguments we conclude that small groups of galaxies most probably
could provide sufficient X-ray emission to reproduce our detection of
extended sources. Accurate data on statistics of groups of galaxies and
their X-ray emission are unavailable. However, the relationship between
various parameters characterizing clusters, such as space density, surface
density of galaxies within a cluster (``richness''), velocity dispersion
and X-ray luminosity show sufficient continuity between rich and poor
clusters (e.g. Edge \& Stewart 1991a,\,b, Henry \& Arnaud 1991, David et al.
1993, Dell'Antonio et al. 1994). Although these data do not provide definitive
estimates of the cluster volume emissivity, we assess that the mean emissivity
of $\sim 5\times 10^{38}$\,erg\,s$^{-1}$Mpc$^{-1}$ indicated by our study is
consistent with the data in the literature. In particular, it coincides with
the luminosity density derived from the luminosity function of the X-ray
selected groups of galaxies obtained by Henry et al. (1995). On the other hand,
the uncertainties of the extrapolation involved in this estimate are large and
we do not claim that the extended emission detected in the present
investigation is definitely and completely produced in poor clusters.

One should note that the temperature of thermal emission by poor groups
of galaxies rarely exceeds 2\,keV (Dell'Antonio et al. 1994).  Thus,
groups of galaxies typically are detected neither by {\em HEAO 1} A-2
nor {\em Ginga} LAC experiments which operate in 2--10\,keV and
4--12\,keV energy bands, respectively. Assuming a thermal emission model,
estimates of the normal galaxy contribution to the XRB above 2\,keV by
Jahoda et al. (1991), Lahav et al. (1993) and Carrera et al. (1995) should
be compared to our calculations which exclude extended sources.
Splitting the total density luminosity into extended
and ``point-like'' emission, the contribution of each component to the XRB
amounts to $\sim 15$\,\%. In this case, the results obtained by ``non-imaging''
instruments of roughly 30\,\% are above our estimates. Taking into account
the large uncertainties of all analyses, this difference is probably
insignificant.

\subsection{Galactic halo}

\ni We now discuss a possibility that extended emission is associated
with individual galaxies rather than groups of galaxies. First, we note
that X-ray halo of $\sim 1$\,Mpc diameter with X-ray luminosity below a
few times $10^{40}$\,erg\,s$^{-1}$ would escape detection as individual
source even in extremely deep X-ray exposures. This is because the surface
brightness of such extended source does not exceed one per cent of the
average XRB intensity and such weak signal falls below the sensitivity
limit in virtually all X-ray experiments. Obviously, a
source with these parameters is not recognized as a single object also in
the present investigation. The correlations analysis could reveal their
existence only if such sources constitute sufficiently numerous class of
X-ray objects.

Normal galaxies have soft X-ray luminosities between $\sim 10^{38}$ to
$\sim 10^{42}$\,erg\,s$^{-1}$ (Fabbiano et al. 1992). Detailed
observations reveal often complex X-ray morphology and quite
frequently there is no one--to--one match between X-ray and optical
features (e.g. Fabbiano 1989, Fabbiano et al. 1992). Although
some galaxies including our own seem to be surrounded by hot gas 
emitting soft X-rays
(e.g. Pietsch \& Tr\"umper 1993, Pietsch 1993, Snowden et al. 1994a,
Ehle et al. 1995), the characteristic size of sources reported in
the previous section would represent a new constituent of
the galaxy X-ray emission. We consider here two alternative models for this
halo. First, galaxies may be surrounded by halos of radius $0.5$\,Mpc of hot
gas which radiates X-rays in thermal bremsstrahlung. The gas
temperature is not constrained by the present measurement. One can
tentatively assume that $kT \la 2$\,keV. In the
second model, the emission is produced by a large number of weak
sources. The possible nature of these sources is unknown, but low mass
X-ray binaries, neutron stars, black holes or subdwarfs are potential
candidates. We refer to a paper by Maoz and Grindlay (1995) who
investigated the possibility of a halo around our Galaxy built with these
objects. They found that all the observational constraints are not
violated if the Galaxy -- and other spirals -- is surrounded by a
population of $\approx 10^{8-9}$ X-ray sources with luminosities
$\approx 10^{30-31}$\,erg\,s$^{-1}$ distributed in a halo with
a characteristic radius of $\sim 15$\,Mpc. It remains to be seen whether
identifications of sources from the deepest {\em ROSAT} pointings
will validate this model or impose observational constrains which
force some modifications of its parameters. In the latter case one
should examine if this model could still provide possible explanation
for the extended emission indicated by the present investigation.

\section{Conclusions and some prospects for the future}

\ni Measurements of the local X-ray volume emissivity are effectively
done using correlations between galaxy distribution and the XRB maps.
Assuming that the X-ray and optical emissivities are proportional when
averaged over large volume of space, we have been able to estimate
total production of soft X-rays per cubic Mpc. Such analysis has been
performed using several galaxy samples with well defined apparent
magnitude limits. Wide range of magnitudes corresponds to substantially
different depths of the galaxy samples. However, because of the broad
distribution of galaxy absolute optical luminosities, galaxies of quite
different apparent magnitudes are spatially correlated. This leads to
significant angular correlations on the celestial sphere between galaxy
samples. In the present analysis all these effects have been accounted for.

X-ray emission associated with the local population of galaxies
contributes to the fluctuations of the XRB on degree scale. The 
amplitude of these fluctuations can be estimated using our present
assessments of the average X-ray luminosity density related to galaxies and
fluctuations of the galaxy distribution. In the next paper of this series
we intend to evaluate the magnitude of the galaxy induced XRB fluctuations
and relate them to the results of Paper I.

The varying ratio of X-ray--to--optical emissivities with redshift and the 
angular dependence of auto- and cross-correlations indicate the presence of
extended X-ray emission. Because of the poor statistics, the intensity
of this diffuse X-ray radiation is not well determined. However, the data
is consistent with the conjecture that about half of the luminosity comes
from regions outside the optical galaxy image. Such extended sources
could be associated with extended emission by hot gas
in groups and clusters of galaxies, although neither our estimates
nor observational data on X-ray properties of such clusters are
sufficiently accurate to make a quantitative comparison.
One could expect even greater observational difficulties
to verify  the ``halo around each galaxy'' model.
If the typical X-ray luminosity of a halo is
$\sim 10^{39}$\,erg\,s$^{-1}$, its detection as a distinct entity would
be practically impossible with the present-day instruments. 
To distinguish between the two models considered here,
one needs extensive statistical data on X-ray
properties of groups of galaxies. We expect also that a scrupulous selection
of ``isolated'' and ``cluster'' galaxies in our samples would help to
address this question.

\acknowledgements
The {\em ROSAT} project has been supported by the Bundesministerium f\"ur
Bildung, Wissenschaft, Forschung und Technologie (BMBF/DARA) and by the
Max-Planck-Society. AMS is grateful to Prof. J.  Tr\"umper for his hospitality
and financial support.  This work has been partially supported by the Polish
KBN grants 2~P304~021~06 and 2~P03D~005~10.


\begin{thebibliography}{}
\bibitem{} 

\bibitem{} Andreani, P., Cristiani, S., 1992, ApJ 398, L13
\bibitem{} Bower, R. G., Hasinger, G., Castander, F. J., et al., 1996, MNRAS
 281, 59
\bibitem{} Boyle, B. J., Griffiths, R. E., Shanks, T., Stewart,
 G. C., Georgantopoulos, I., 1993, MNRAS 260, 49
\bibitem{} Briel, U. G.,  Henry, J. P., 1993, A\&A 278, 390
\bibitem{} Butcher, H. R., van Breugel, W., Miley, G. K., 1980, ApJ 235, 749
\bibitem{} Carrera, F. J., Barcons, X., Butcher, J. A., et al., 1995, MNRAS
 275, 22
\bibitem{} David, L. P., Slyz, A., Jones, C., et al., 1993, ApJ 412, 479
\bibitem{} Dell'Antonio, I. P., Geller, M. J., Fabricant, D. G., 1994, AJ
 107, 427
\bibitem{} Driver, S. P., Phillipps, S., Davies, J. I., Morgan, I., Disney, M.
 J., 1994, MNRAS 266, 155
\bibitem{} Ebeling, H., 1993, Abell and ACO Clusters of Galaxies in the ROSAT
 All-Sky X-ray Survey: A Statistical Study. PhD Thesis, Garching, MPE report 250
\bibitem{} Ebeling, H., Allen, S. W., Crawford, C. S., et al., 1996, In:
 Zimmermann, H. U., Tr\"umper, J. (eds.) R\"ontgenstrahlung from the Universe,
 Garching, MPE report 263, p. 579 
\bibitem{} Edge, A. C., Stewart, G. C., Fabian, A. C., Arnaud, K. A., 1990,
 MNRAS 245, 559
\bibitem{} Edge, A. C., Stewart, G. C., 1991a, MNRAS 252, 414
\bibitem{} Edge, A. C., Stewart, G. C., 1991b, MNRAS 252, 428
\bibitem{} Ehle, M., Pietsch, W., Beck, R., 1995, A\&A 295, 289
\bibitem{} Elvis, M., Soltan, A., Keel, W. C., 1984, ApJ 283, 479
\bibitem{} Eskride, P. B., Fabbiano, G., Kim, D.-W., 1995, ApJS 97, 141
\bibitem{} Fabbiano, G., 1989, ARA\&A 27, 87
\bibitem{} Fabbiano, G., Kim, D.-W., Trinchieri, G., 1992, ApJS 80, 531
\bibitem{} Hasinger, G., 1992, {\em ROSAT} Deep Surveys. In: Brinkmann, W.,
 Tr\"umper, J. (eds.) Proc. MPE Conf., X-ray Emission from Active Galactic
 Nuclei and the Cosmic X-ray Background. Garching, MPE report 235, p. 321
\bibitem{} Hasinger, G., Burg, R., Giacconi, R., et al., 1993, A\&A 275, 1
 (Erratum: A\&A 291, 348)
\bibitem{} Henry, J. P., Arnaud, K. A., 1991, ApJ 372, 410
\bibitem{} Henry, J. P., Gioia, I. M., Maccacaro, T., et al., 1992, ApJ
 386, 408
\bibitem{} Henry, J. P., Gioia, I. M., Huchra, J. P., et al., 1995, ApJ
 449, 422
\bibitem{} Iovino, A., Shaver, P. A., 1988, ApJ 330, L13
\bibitem{} Jahoda, K., Lahav, O., Mushotzky, R. F., Boldt, E., 1991, ApJ
 378, L37
\bibitem{} Jahoda, K., Lahav, O., Mushotzky, R. F., Boldt, E., 1992, ApJ
 399, L107
\bibitem{} Kruszewski, A., 1988, Acta Astron., 38, 155
\bibitem{} Lahav, O., Fabian, A. C., Barcons, X., et al., 1993, Nat 364, 693
\bibitem{} Lauberts, A., 1982, The ESO/Uppsala Survey of the ESO(B) Atlas,
 ESO, Garching
\bibitem{} Maccacaro, T., Gioia, I. M., Wolter, A., Zamorani, G., Stocke,
 J. T., 1988, ApJ 326, 680
\bibitem{} Maoz, E., Grindlay, J., 1995, ApJ 444, 183
\bibitem{} Miyaji, T., Lahav, O., Jahoda, K., Boldt, E., 1994, ApJ 434, 424
\bibitem{} Mo, H. J., Fang, L. Z., 1993, ApJ 410, 493
\bibitem{} Nilson, P., 1973, Uppsala General Catalogue of Galaxies, Uppsala
 Astr. Obs. Ann. Vol. 6
\bibitem{} Paturel, G., Fouqu\'e, P., Bottinelli, L., Gouguenheim, L., 1989,
 A\&AS 80, 299
\bibitem{} Peebles, P. J. E., 1980, The Large Scale
 Structure of the Universe. Princeton Univ. Press, Princeton, p. 155
\bibitem{} Pfeffermann, E., et al., 1987, In: Koch, E.-E., Schamhl, G. (eds.)
 Soft X-Ray Optics and Technology (Proc. SPIE 733, 519)
\bibitem{} Pietsch, W., 1993, In: Hensler, G., Theis, C., Gallagher, J. (eds.)
 Panchromatic View of Galaxies, Gif-sur-Yvette Cedex, Edition Frontiers, p. 137
\bibitem{} Pietsch, W., Tr\"umper, J., 1993, Adv. Space Res. 13, 171
\bibitem{} Roche, N., Shanks, T., Georgantopoulos, I., et al., 1995, MNRAS
 273, L15
\bibitem{} Schechter, P., 1976, ApJ 203, 297
\bibitem{} Seldner, M., Peebles, P. J. E., 1977, ApJ 215, 703
\bibitem{} Shane, C. D., Wirtanen, C. A., 1967, Publ. Lick Obs.,
 Vol. XXII--part I
\bibitem{} Snowden, S. L., Freyberg, M. J., Plucinsky, P. P., et al., 1995, ApJ
 454, 643
\bibitem{} Snowden, S. L., Hasinger, G., Jahoda, K., et al., 1994a, ApJ
 430, 601
\bibitem{} Snowden, S. L., McCammon, D., Burrows, D., Mendenhall, J. A., 1994b,
 ApJ 424, 714
\bibitem{} Snowden S. L., Schmitt, J. H. M. M., 1990, Ap\&SS 171, 207
\bibitem{} So\l tan, A., Hasinger, G., 1994, A\&A 288, 77
\bibitem{} So\l tan, A. M., Hasinger, G., Egger, R., Snowden, S.,
 Tr\"umper, J., 1996, A\&A 305, 17 (Paper I)
\bibitem{} Treyer, M. A., Lahav, O., 1995, MNRAS submitted, SISSA bulletin
 board, astro-ph/9509013
\bibitem{} Tr\"umper, J., 1983, Adv. Space Res. 4, (4)241
\bibitem{} Turner, E. L., Geller, M. J., 1980, ApJ 236, 1
\bibitem{} Voges, W., 1992, In: Guyenne, T. D., Hunt, J. J. (eds.) Science with
 particular emphasis on High-Energy Astrophysics, Proc. of Satellite
 Symposium 3, Space (Noordwijk: ESA Publication Division), 9
\end{thebibliography}
\end{document}